\documentstyle[12pt]{article}

\footnotetext[1]{Corresponding author.\\ E-mail address: wq-tj@china.com}
\begin{document}
\title{Effect of quantum lattice fluctuations on the optical-absorption
spectra of halogen-bridged mixed-valence transition-metal complexes}
\author{Q. Wang$^{{\rm a}}${\footnotemark[1] } \ \ \ \ and \ \ \ \
H. Zheng$^{{\rm b}}$ \\
%EndAName
{\normalsize {\em $^{{\rm a,b}}$Department of Applied Physics, Shanghai Jiao
Tong University,}}\\
{\normalsize {\em Shanghai 200030, People's Republic of China}}\\
{\normalsize {\em $^{{\rm a}}$Physics Department, Tongji University,
Shanghai 200092, People's Republic of China}}\\
{\normalsize {\em $^{{\rm a}}$Physics Department, Xinjiang Normal
University, Uromuq 830053, People's Republic of China}}}
\date{}
\maketitle

\begin{abstract}
The effect of quantum lattice fluctuations on the optical-absorption spectra
in the ground state of halogen-bridged mixed-valence transition-metal
linear-chain complexes is studied by using a one-dimensional extended
Peierls-Hubbard model. The nonadiabatic effects due to finite phonon
frequency $\omega_{\pi}>0$ are treated through an energy-dependent
electron-phonon scattering function $\delta(k^{\prime},k)$ introduced by
means of an unitary transformation. The calculated optical-absorption
spectra do not have the inverse-square-root singularity, but they have a
peak above the gap edge and there exists a significant tail below the peak,
which are consistent with the optical-absorption coefficient or the optical
conductivity measurements of this material.\newline
\end{abstract}

\newpage

\section{Introduction}

Halogen-bridged mixed-valence transition-metal complexes (HMMC's or $MX$
chain), as one of typical materials with a quasi-one-dimensional
charge-density-wave (CDW) state, have been the subject of intense studies in
recent years for both theoretists and experimentalists, because of the
intrinsically interesting properties and the important technological
applications of the materials \cite{refb1,refa1,refa2}. Each chain of the
materials is composed of transition-metal ions $M^{+3}$ (=Pt$^{+3}$,
Pd$^{+3}$, Ni$^{+3}$) bridged by halogen ions $X^{-}$ (=Cl$^{-}$, Br$^{-}$,
I$^{-}$), with ligands attached to the metals and in some cases counterions
between the chains to maintain charge neutrality. $M^{+3}$ has an unpaired
electron in its $d_{z^{2}}$ orbital (\^{\bf z} is parallel to the chain), and
this orbital makes an energy band through the supertransfer between
neighbouring
two $d_{z^{2}}$ orbitals. This supertransfer comes from the hybridization
between the $d_{z^{2}} $ orbital and the $p_{z} $ orbital of $X $. Because
of the electron-phonon (e-ph) coupling, the charge transfer occurs between
neighbouring two $M^{\prime}s $ so as to give the mixed-valence state. This
is nothing but the CDW state with twice the period of the original lattice.
The $MX$ chain materials constitute a rapidly expending class of near
single-crystal quasi-one-dimensional systems with phases of ground states
ranging from strong CDW to spin-density-wave or spin-Peierls \cite{refa3}.

Many previous investigations of $MX$ chain materials used the adiabatic
approximation for the phonons. The adiabatic approximation means neglect of
the dynamic energy of halogen ions, that is, neglect of the effect of
lattice vibration. In nonadiabatic case, the phonon frequency $\omega_{\pi}$
is finite. Generally speaking, the nonadiabatic effect will suppress the
state order parameters of the system \cite{refa4,refa5}. As far as the
optical-absorption coefficient is concerned, the results of adiabatic
approximation methods have inverse-square-root singularity at the gap edge.
By considering the nonadiabatic effect, the inverse-square-root singularity
at the gap edge may disappear \cite{refa6}. The relationship between the
change of optic-absorption shape and the phonon frequency in the range from
$\omega_{\pi}=0$ to $\omega_{\pi}\rightarrow\infty$ should be studied for
understanding the physical properties of e-ph interactions of $MX$ chain in
nonadiabatic case.

The one-dimensional one-band model for $MX$ chain complexes can be written
as \cite{refa7,refa8,refa9} 
\begin{equation}
H=\sum_{l}\left(\frac{1}{2M}P^{2}_{l}+\frac{K}{2}u^{2}_{l}\right)
-\sum_{l,s}t_{0}(c^{\dag}_{l,s}c_{l+1,s}+c^{\dag}_{l+1,s}c_{l,s})
+\sum_{l,s}\alpha(u_{l}-u_{l+1})c^{\dag}_{l,s}c_{l,s}+H_{c},
\label{eq:hamilt}
\end{equation}
where 
\begin{equation}
H_{c}=U\sum_{l}n_{l,\uparrow}n_{l,\downarrow}+V\sum_{l,s,s^{\prime}}
n_{l,s}n_{l+1,s^{\prime}}.
\end{equation}
Here $c^{\dag}_{l,s}$ and $c_{l,s}$ are the creation and annihilation
operators of electrons sitting at site $l$ with spin $s$, $u_{l}$ (it's
conjugated momentum is $p_{l}$) is the displacement of the $l$th ion, $t_{0}$
is the supertransfer energy of electrons between neighbouring two orbitals,
$\alpha$ is the e-ph coupling constant, $K$ is the elastic constant and $M$
the mass of ions. $U$ and $V$ correspond to the on-site and the
nearest-neighbour Coulomb energies, respectively (throughout this paper,
we set $\hbar=k_{B}=1$).

\section{Theoretical analysis}

In the Hamiltonian (\ref{eq:hamilt}), the operators of lattice modes, $u_{l}$
and $p_{l}$, can be expanded by the phonon creation and annihilation
operators $b^{\dag}_{-q}$ and $b_{q}$, and after Fourier transformation, the 
$H$ reads 
\begin{eqnarray}
H&=&H^{0}+H_{ep}+H_{c},
\end{eqnarray}
where 
\begin{eqnarray}
H^{0}&=&\sum_{q}\omega_{\pi}\left(b^{\dag}_{q}b_{q}+\frac{1}{2}\right) +
\sum_{k,s}\epsilon_{k}c^{\dag}_{k,s}c_{k,s}, \\
H_{ep}&=&\sum_{q,k,s}\frac{1}{\sqrt{N}}g(q)(b_{q}+b^{\dag}_{-q})
c^{\dag}_{k+q,s}c_{k,s}.
\end{eqnarray}
Here, $\epsilon_{k}=-2t_{0}\cos k$ is the bare band function, and $N$ is the
total number of sites. The coupling function
$g(q)=\alpha\sqrt{1/(2M\omega_{\pi})}[1-\exp(iq)]$ and the dispersionless
phonon frequency $\omega_{\pi}=\sqrt{K/M}$.

In order to take into account the fermion-phonon correlation, an unitary
transformation is applied to $H$ \cite{refa12,refa13}, 
\begin{equation}
H^{\prime}=\exp(S)H\exp(-S),
\end{equation}
where the generator $S$ is 
\begin{equation}
S=\frac{1}{\sqrt{N}}\sum_{q,k,s}\frac{g(q)}{\omega_{\pi}} (b^{%
\dag}_{-q}-b_{q})\delta(k+q,k)c^{\dag}_{k+q,s}c_{k,s}.
\end{equation}
Here we introduce a function $\delta(k^{^{\prime}},k)$ which is a function
of the energies of the incoming and outgoing fermions in the fermion-phonon
scattering process and will be determined later. The unitary transformation
can proceed order by order, 
\begin{equation}
H^{\prime}=H^{0}+H_{ep}+H_{c}+[S,H^{0}]+[S,H_{ep}]+[S,H_{c}]
 +\frac{1}{2}[S,[S,H^{0}]]+\frac{1}{2}[S,[S,H_{c}]]+O(\alpha^{3}).
\end{equation}
The first-order terms in $H^{\prime}$ are 
\begin{eqnarray}
H_{ep}+[S,H^{0}] & = & \frac{1}{\sqrt{N}}\sum_{q,k,s}g(q)(b^{\dag}_{-q}
      +b_{q}) c^{\dag}_{k+q,s}c_{k,s}  \nonumber \\
& & -\frac{1}{\sqrt{N}}\sum_{q,k,s}g(q)\delta(k+q,k)
(b^{\dag}_{-q}+b_{q})c^{\dag}_{k+q,s}c_{k,s}  \nonumber \\
& & +\frac{1}{\sqrt{N}}\sum_{q,k,s} \frac{g(q)}{\omega_{\pi}}\delta(k+q,k)
(b^{\dag}_{-q}-b_{q})(\epsilon_{k}-\epsilon_{k+q}) c^{\dag}_{k+q,s}c_{k,s}.
\end{eqnarray}
Note that the ground state $|g_{0}\left.\right\rangle$ of $H^{0}$, the
non-interacting system, is a direct product of a filled fermi-sea
$|FS\rangle$ and a phonon vacuum state $|ph,0\left.\right\rangle$
\cite{refa14}:
\begin{equation}
|g_{0}\left.\right\rangle=|FS\left.\right\rangle|ph,0\left.\right\rangle.
\end{equation}
Applying the first-order terms on $|g_{0}\left.\right\rangle$ we get 
\begin{equation}
(H_{ep}+[S,H^{0}])|g_{0}\left.\right\rangle
 =\frac{1}{\sqrt{N}}\sum_{q,k,s}g(q)b^{\dag}_{-q}
 c^{\dag}_{k+q,s}c_{k,s}\left[1-\delta(k+q,k)\left(1-
 \frac{\epsilon_{k}-\epsilon_{k+q}}{\omega_{\pi}}\right)\right]
 |g_{0}\left.\right\rangle,
\end{equation}
since $b_{q}|ph,0\left.\right\rangle=0$. As the band is half-filled, the
Fermi energy $\epsilon_{F}=0$. Thus
$c^{\dag}_{k+q}c_{k}|FS\left.\right\rangle\not=0$ only if
$\epsilon_{k+q}\ge 0$ and $\epsilon_{k}\le 0$. So, we have 
\begin{equation}
(H_{ep}+[S,H^{0}])|g_{0}\left.\right\rangle=0,
\end{equation}
if we choose 
\begin{equation}
\delta(k+q,k)=1/(1+|\epsilon_{k+q}-\epsilon_{k}|/\omega_{\pi}).
\end{equation}
This is nothing but making the matrix element of $H_{ep}+[S,H^{0}]$ between
$|g_{0}\left.\right\rangle$ and the lowest-lying excited states vanishing.
Thus the first-order terms, which are not exactly canceled after the
transformation, are related to the higher-lying excited states and should be
irrelevant under renormalization \cite{refa14}. The second-order terms in
$H^{\prime}$ can be collected as follows: 
\begin{eqnarray}
& & [S,H_{ep}]+\frac{1}{2}[S,[S,H^{0}]]  \nonumber \\
& = & \frac{1}{2N}\sum_{q,k,s}\sum_{q^{\prime},k^{\prime}}
\frac{g(q)g(q^{\prime})}{\omega_{\pi}} \delta(k+q,k)[2-\delta(k^{\prime}
+q^{\prime},k^{\prime})](b^{\dag}_{-q^{\prime}}+b_{q^{\prime}})
(b^{\dag}_{-q}-b_{q})  \nonumber \\
& & \times(c^{\dag}_{k+q,s}c_{k^{\prime},s}\delta_{k,k^{\prime}+q^{\prime}}
-c^{\dag}_{k^{\prime}+q^{\prime},s}c_{k,s}\delta_{k^{\prime},k+q})  \nonumber
\\
& & +\frac{1}{2N}\sum_{q,k,s}\sum_{q^{\prime},k^{\prime}}
\frac{g(q)g(q^{\prime})} {\omega_{\pi}^{2}}\delta(k+q,k)\delta(k^{\prime}
+q^{\prime},k^{\prime})(\epsilon_{k^{\prime}}-
\epsilon_{k^{\prime}+q^{\prime}})(b^{\dag}_{-q^{\prime}}-b_{q^{\prime}})
(b^{\dag}_{-q}-b_{q})  \nonumber \\
& & \times(c^{\dag}_{k+q,s}c_{k^{\prime},s}\delta_{k,k^{\prime}+q^{\prime}}
-c^{\dag}_{k^{\prime}+q^{\prime},s}c_{k,s}\delta_{k^{\prime},k+q})  \nonumber
\\
& & -\frac{1}{2N}\sum_{q,k,s}\sum_{k^{\prime},s^{\prime}}
\frac{g(q)g(-q)}{\omega_{\pi}} \delta(k+q,k)
[2-\delta(k^{\prime}-q,k^{\prime})]c^{\dag}_{k+q,s}c_{k,s}
c^{\dag}_{k^{\prime}-q,s^{\prime}}c_{k^{\prime},s^{\prime}}.
\end{eqnarray}
$\delta_{k^{\prime},k+q}$ is the Kronecker $\delta$ symbol. All terms of
higher order than $\alpha^{2}$ will be omitted in the following treatment.
Note that the first order terms of transformed $H_{c}$ have not been
included in the process of determining function $\delta(k^{\prime},k)$ (from
Eq. (9) to Eq. (13)) and the second order terms of that have not been
included in Eq. (14). The reason for doing so will be discussed later.

Then we make a displacement transformation to $H^{\prime}$ to take into
account the static phonon-staggered ordering \cite{refa6}, 
\begin{equation}
\tilde{H}=\exp(R)H^{\prime}\exp(-R).
\end{equation}
Here, 
\begin{equation}
R=-\sum_{l}(-1)^{l}u_{0}\sqrt{\frac{M\omega_{\pi}}{2}}(b^{\dag}_{l}-b_{l}),
\end{equation}
and $\exp(R)$ is a displacement operator.

If the ground state of $H$ is $|g\rangle$, then the ground state of
$\tilde{H}$ is $|g^{\prime}\rangle$:
$|g\rangle=\exp(-S)\exp(-R)|g^{\prime}\rangle$.
We assume that for $|g^{\prime}\rangle$, the fermions and phonons can be
decoupled: $|g^{\prime}\rangle\approx |fe\rangle|ph,0\rangle$, where
$|fe\rangle$ is the ground state for fermions. After averaging $\tilde H$
over the phonon vacuum state, we get an effective Hamiltonian for the
fermions, 
\begin{eqnarray}
H_{{\rm eff}} & = & \left\langle ph,0|\tilde{H}|ph,0\right\rangle \nonumber\\
& = & \frac{1}{2}KNu_{0}^{2}+H_{c} +\sum_{k,s}E_{0}(k)c^{\dag}_{k,s}c_{k,s}
+\sum_{k>0,s}\Delta_{0}(k)(c^{\dag}_{k-\pi,s}c_{k,s}
+c^{\dag}_{k,s}c_{k-\pi,s})  \nonumber \\
& & -\frac{1}{N}\sum_{q,k,s}\sum_{k^{\prime},s^{\prime}} \left\{\frac{%
g(q)g(-q)}{\omega_{\pi}} \delta(k+q,k)[2-\delta(k^{\prime}-q,k^{\prime})]
-V_{0}\delta_{s,s^{\prime}}\right\}  \nonumber \\
& & \times c^{\dag}_{k+q,s}c_{k,s}c^{\dag}_{k^{\prime}-q,s^{\prime}}
c_{k^{\prime},s^{\prime}}.
\end{eqnarray}
Where 
\begin{eqnarray}
E_{0}(k)&=&\epsilon_{k}-\frac{1}{N}\sum_{k^{\prime}}
\frac{g(k^{\prime}-k)g(k-k^{\prime})}{\omega^{2}_{\pi}}\delta(k^{\prime},k)
\delta(k,k^{\prime}) (\epsilon_{k}-\epsilon_{k^{\prime}}), \\
\Delta_{0}(k)&=&2\alpha u_{0}[1-\delta(k-\pi,k)], \\
V_{0}&=&\frac{1}{N^{3}}\sum_{k,k^{\prime},q} \frac{g(q)g(-q)}{\omega_{\pi}}
\delta(k+q,k)[2-\delta(k^{\prime}-q,k^{\prime})].
\end{eqnarray}

Now, the total Hamiltonian can be divided as
$\tilde{H}=\tilde{H}_{0}+\tilde{H}_{1}$, where $\tilde{H}_{1}$ includes the
terms that are zero after being averaged over the phonon vacuum state, and
\begin{equation}
\tilde{H}_{0}=\sum_{q}\omega_{\pi}\left(b^{\dag}_{q}b_{q}+\frac{1}{2}\right)
+H_{{\rm eff}}.
\end{equation}
After the unitary transformations, the Coulomb interaction $H_{c}$ becomes
$H_{c}+$higher order terms. The contributions of these higher order terms to
$\tilde{H}_{0}$ are zero under the mean-field approximation. So, in the
process of determining function $\delta(k^{\prime},k))$, which only concerns
the first order terms of transformed Hamiltonian, and also in Eq. (14) and
Eq. (17), we have not written out the higher order terms of transformed
$H_{c}$. By using the mean-field approximation to the terms of four-fermion
interaction in $H_{{\rm eff}}$, we have 
\begin{equation}
\tilde{H}_{0}=\frac{1}{2}KNu_{0}^{2}+
\sum_{q}\omega_{\pi}\left(b^{\dag}_{q}b_{q}+
\frac{1}{2}\right)+\sum_{k,s}E_{k}c^{\dag}_{k,s}c_{k,s}
+\sum_{k>0,s}\Delta_{k}(c^{\dag}_{k-\pi,s}c_{k,s}
+c^{\dag}_{k,s}c_{k-\pi,s}).
\end{equation}
In the theoretical analysis, we have distinguished different physical
processes and taken into account the fact that only the Umklapp scattering
terms affect the gap, and the forward and backward scattering terms
contribute nothing to the "charge" gap \cite{refa17,refa18}. The
renormalized band function and the gap function are 
\begin{eqnarray}
E_{k}&=&E_{0}(k)-\frac{2}{N}\sum_{k^{\prime}}\left\{\frac{\alpha^{2}}{K}
\sin^{2}(\frac{k^{\prime}-k}{2})\delta(k^{\prime},k)
[2-\delta(k^{\prime},k)]-V\cos(k^{\prime}-k)\right\}
\frac{E_{k^{\prime}}}{W_{k^{\prime}}}, \\
\Delta_{k}&=&2\alpha u_{0}[c-\frac{1}{2}\delta(k-\pi,k)].
\label{delta}
\end{eqnarray}
Where 
\begin{equation}
c=1+\frac{{1}}{N}\sum_{k>0}\left[2V-\frac{1}{2}U-V_{0}+ \frac{4\alpha^{2}}{K}%
\delta(k-\pi,k)\right] \frac{\Delta_{k}}{2\alpha u_{0}W_{k}},
\end{equation}
and $W_{k}=\sqrt{E^{2}_{k}+\Delta^{2}_{k}}$ is the fermionic spectrum of
elementary excitations in the dimerized state.

We find by means of the variational principle that the equation to determine
dimerized lattice displacement ordering parameter $u_{0}$ is 
\begin{equation}
1=\frac{4\alpha^{2}}{KN}\sum_{k>0,s}[1-\delta(k-\pi,k)] \frac{\Delta_{k}}{%
2\alpha u_{0}W_{k}}.
\label{uofun}
\end{equation}
If $\omega_{\pi}=0$ we have $\delta(k^{\prime},k)=0$. Eq. (\ref{uofun})
becomes the same as that in the adiabatic theory. In our theory
$\delta(k-\pi,k)$ has a sharp peak at the Fermi point and, since 
\begin{equation}
1-\delta(k-\pi,k)=\frac{4t_{0}|\cos k|}{\omega_{\pi}+4t_{0}|\cos k|},
\end{equation}
the logarithmic singularity in the integration of Eq. (\ref{uofun}) in the
adiabatic case is removed by the factor $1-\delta(k-\pi,k)$ as long as the
ratio $\omega_{\pi}/t_{0}$ is finite. For the gapped state (insulating
state) we believe that the mean-field approximation is still an effective
one. The reason is as follows: The quadratic part of $H_{{\rm eff}}$ (which
can be exactly diagonalized) contains a gap term ($\Delta_{0}(k)\neq 0$)
that is related to the Peierls dimerization and suppresses the quantum
fluctuations. Furthermore, the logarithmic singularity (related to the
infrared divergence of low dimensional system) has been removed by the
factor $1-\delta(k-\pi,k)=4t_{0}|\cos k|/(\omega_{\pi}+4t_{0}|\cos k|)$.
Comparing Eq. (\ref{delta}) with that in the adiabatic case,
$\Delta=2\alpha u_{0}$, we have the gap in the nonadiabatic case,
$\Delta=\Delta(\pi/2)=2\alpha u_{0}[c-1/2]$, which is nonzero but smaller
than the adiabatic gap. This is the true gap in the fermionic spectrum.

\section{Optical absorption}

The optical-absorption coefficient $\alpha(\omega)$ can be expressed by the
retarded Green's function as follows: 
\begin{equation}
\alpha(\omega)=-\frac{2}{\pi\omega}{\rm Im}K^{R}(\omega),
\end{equation}
where $K^{R}$ is defined as 
\begin{equation}
K^{R}(\omega)=-i\int^{0}_{-\infty}e^{-i\omega t}dt \langle
g|[j(0)j(t)-j(t)j(0)]|g\rangle.
\end{equation}
Here, $j$ is the current operator \cite{refa8}, 
\begin{eqnarray}
j & = & -iet_{0}\sum_{l,s}(c^{\dag}_{l,s}c_{l+1,s}-c^{\dag}_{l+1,s}c_{l,s}) 
\nonumber \\
& = & J\sum_{k,s}\sin k c^{\dag}_{k,s}c_{k,s},
\end{eqnarray}
where $J=2et_{0}$, and $j(t)=\exp(iHt)j\exp(-iHt)$ is the form of $j$ in the
Heisenberg representation. As $R$ commutes with $j$, the unitary
transformation of the current operator is 
\begin{eqnarray}
e^{S}je^{-S}&=&j+[S,j]+\frac{1}{2}[S,[S,j]]+O(\alpha^{3})  \nonumber \\
& = & J\sum_{k>0,s}\sin k(c^{\dag}_{k,s}c_{k,s}
-c^{\dag}_{k-\pi,s}c_{k-\pi,s})  \nonumber \\
& & +\frac{J}{\sqrt{N}}\sum_{k>0,q,s}\frac{g(q)}{\omega_{\pi}}
(b^{\dag}_{-q}-b_{q})\delta(k+q,k)[\sin k-\sin(k+q)]
(c^{\dag}_{k+q,s}c_{k,s}-c^{\dag}_{k+q-\pi,s}c_{k-\pi,s})  \nonumber \\
& & +\frac{J}{2N}\sum_{k>0,q,q^{\prime},s}
\frac{g(q)g(q^{\prime})}{\omega_{\pi}^{2}}
(b^{\dag}_{-q}-b_{q})(b^{\dag}_{-q^{\prime}}-b_{q^{\prime}})\delta(k+q,k)
[\sin k-\sin(k+q)]  \nonumber \\
& &\times[\delta(k+q+q^{\prime},k+q)(c^{\dag}_{k+q+q^{\prime},s}c_{k,s}
-c^{\dag}_{k+q+q^{\prime}-\pi,s}c_{k-\pi,s})  \nonumber \\
& & -\delta(k,k-q^{\prime})(c^{\dag}_{k+q,s}c_{k-q^{\prime},s}-
c^{\dag}_{k+q-\pi,s}c_{k-q^{\prime}-\pi,s})].
\end{eqnarray}
Because the averaging of $\tilde{H_{1}}$ over the phonon vacuum state is
zero, in the ground state at zero temperature, $\tilde{H_{1}}$ can be
neglected. By using the approximately decoupling
$|g^{\prime}\rangle\approx|g_{0}^{\prime}\rangle$, the ground state of
$\tilde{H}_{0}$, and $\tilde{H}\approx\tilde{H}_{0}$ \cite{refa14} in the
calculation
\begin{eqnarray}
\langle g|j(0)j(t)|g\rangle & = & \langle g^{\prime}|[e^{(S+R)}je^{-(S+R)}]
e^{i\tilde{H}t}[e^{(S+R)}je^{-(S+R)}]e^{-i\tilde{H}t}|g^{\prime}\rangle 
\nonumber \\
& \approx & \langle g^{\prime}_{0}|[e^{S}je^{-S}]e^{i\tilde{H}_{0}t}
[e^{S}je^{-S}]e^{-i\tilde{H}_{0}t}|g^{\prime}_{0}\rangle,
\end{eqnarray}
we can get 
\begin{eqnarray}
K^{R}(\omega)&=&J^{2}\sum_{k>0,s}\left(\frac{1}{\omega-2W_{k}+i0^{+}}
-\frac{1}{\omega+2W_{k}-i0^{+}}\right)  \nonumber \\
& & \times\left\{\sin^{2}k+\frac{2\alpha^{2}}{KN\omega_{\pi}}
\sum_{k^{\prime}}[1-\cos(k^{\prime}-k)]\delta^{2}(k^{\prime},k) \sin k(\sin
k^{\prime}-\sin k)\right\}\frac{\Delta^{2}_{k}} {W^{2}_{k}}  \nonumber \\
& & +\frac{\alpha^{2}J^{2}}{KN^{2}\omega_{\pi}} \sum_{k>0,k^{\prime}>0,s}
\left (\frac{1}{\omega-\omega_{\pi}-W_{k}-W_{k^{\prime}}+i0^{+}}
-\frac{1}{\omega+\omega_{\pi}+W_{k}+W_{k^{\prime}}-i0^{+}}\right) \nonumber \\
& & \times\left\{[1-\cos(k^{\prime}-k)]\delta^{2}(k^{\prime},k)(\sin
k^{\prime}-\sin k)^{2}(\alpha_{k}\beta_{k^{\prime}}
+\beta_{k}\alpha_{k^{\prime}})^{2}\right.  \nonumber \\
& & +\left.[1+\cos(k^{\prime}-k)]\delta^{2}(k^{\prime}-\pi,k)(\sin
k^{\prime}+\sin k)^{2}(\alpha_{k}\alpha_{k^{\prime}}
+\beta_{k}\beta_{k^{\prime}})^{2}\right\},
\end{eqnarray}
where $\alpha_{k}=\sqrt{(1+E_{k}/W_{k})/2}$, and
$\beta_{k}=\sqrt{(1-E_{k}/W_{k})/2}$. Thus, we have the
optical-absorption coefficient
\begin{eqnarray}
\alpha(\omega) & = &\frac{2J^{2}(2s+1)}{\omega N}\sum_{k>0}
\delta(\omega-2W_{k})  \nonumber \\
& &\times\left\{\sin^{2}k+ \frac{2\alpha^{2}}{KN\omega_{\pi}}
\sum_{k^{\prime}}[1-\cos(k^{\prime}-k)]\delta^{2}(k^{\prime},k) \sin k(\sin
k^{\prime}-\sin k)\right\}\frac{\Delta^{2}_{k}} {W^{2}_{k}}  \nonumber \\
& &+\frac{2\alpha^{2}J^{2}(2s+1)}{\omega KN^{2}\omega_{\pi}}
\sum_{k>0,k^{\prime}>0} \delta(\omega-\omega_{\pi}-W_{k}-W_{k^{\prime}}) 
\nonumber \\
& &\times\left\{[1-\cos(k^{\prime}-k)]\delta^{2}(k^{\prime},k) (\sin
k^{\prime}-\sin k)^{2}(\alpha_{k}\beta_{k^{\prime}}
+\beta_{k}\alpha_{k^{\prime}})^{2}\right.  \nonumber \\
& &+\left.[1+\cos(k^{\prime}-k)]\delta^{2}(k^{\prime}-\pi,k) (\sin
k^{\prime}+\sin k)^{2}(\alpha_{k}\alpha_{k^{\prime}}
+\beta_{k}\beta_{k^{\prime}})^{2}\right\}.
\end{eqnarray}

The relationships between the change of the optical-absorption shape and the
different phonon frequencies are shown in Figure 1. The parameter values
used are: $\alpha^{2}/K=0.5t_{0}$, $U=1.0t_{0}$ and $V=0.15t_{0}$ with
$\omega_{\pi}=0.004t_{0}, 0.008t_{0}$, and $0.012t_{0}$. One can see that the
spectrum broadens but the peak height decreases as $\omega_{\pi}$ increases.
The inverse-square-root singularity at the gap edge in the adiabatic case
\cite{refa19} disappears because of the nonadiabatic effect. For comparison,
the adiabatic $(\omega_{\pi}=0)$ result is also shown which has an
inverse-square-root singularity at gap edge. For finite $\omega_{\pi}$, the
singularity is absent and there is a significant tail below the peak. We
note that in our theory, in mathematical viewpoint, the difference between
the $\omega_{\pi}=0$ and $\omega_{\pi}>0$ cases mainly comes from the
functional form of the gap [see Eq. (\ref{delta})]. Comparing it with that
in the adiabatic limit, one can see that the subgap states come from the
quantum lattice fluctuations, i.e., the second term in the square bracket of
Eq. (\ref{delta}).

Figure 2 shows the calculated optical-absorption coefficients
$\alpha(\omega)$ versus the normalized photon frequency relations for
$\omega_{\pi}=0.008t_{0}$, $U=1.0t_{0}$ and $V=0.15t_{0}$ in the cases of
$\alpha^{2}/K=0.3t_{0}$, $0.4t_{0},$ and $0.5t_{0}$. One can see that as the
e-ph coupling constant $\alpha^{2}/K$ increases, the peak of
optical-absorption spectrum moves to higher photon energy and the
dimerization gap becomes wider.

Figure 3 (a) and Figure 3 (b) show the calculated optical-absorption
coefficients $\alpha(\omega)$ versus the normalized photon frequency
relations for $\omega_{\pi}=0.008t_{0}$ and $\alpha^{2}/K=0.5t_{0}$ in the
cases of different $U$ and $V$ values, respectively. One can see that the
spectra are generally the same, except that the peak height decreases and
shifts to higher photon energy when $U$ decreases or $V$ increases. The
on-site and the nearest-neighbour Coulomb repulsions are to weaken and
strengthen the lattice dimerization, respectively. In the $MX$ chain, a
charge-transfer (CT) excitation originates in the electron excitation over
the CDW gap. If no definite fine structure can be observed in the absorption
band, it cannot be clarified, only from the line shape of
optical-absorption, whether this band is due to the CT exciton or the
interband transition. Wada {\it et al.} \cite{refc1} claimed that because of
the electric field dependence of the modulation signal observed in the
measurement of the electroreflectance spectra of the $MX$ chain, it can be
excluded that the CT excitation absorption band is an interband absorption.
However, K. Iwano {\it et al}. \cite{refc2} indicated that in the case of
weak electron-hole binding, i.e., small $V$, the spectrum becomes the
mixture of the exciton absorption and the interband ones. If the value of $V$
is reduced toward zero, the exciton contribution then becomes more and more
small and, finally, the exciton disappears and the spectrum is now given
completely by the interband transitions. Our calculated result of the $V=0$
case is also shown in Fig. 3 (b) and this is the optical-absorption spectrum
of the interband transitions.

In Figure 4, we compare our calculation (solid line) with the observed
optical absorption (solid circles) of $MX$ chain complexes \cite{refc3}.
Here we use the same input parameters (without adjusting them) $U=1.0t_{0}$
and $V=0.3t_{0}$, as those of K. Iwano and K. Nasu \cite{refc2}. Other input
parameters used are $\omega _{\pi }=0.01t_{0}$ and $\alpha ^{2}/K=0.2t_{0}$.
One can see that the agreement between the experiment result and our
calculation is quite good. In experiments, the measurement of
optical-absorption spectrum could be affected by various
factors, such as the impurity in samples, the finite
measurement temperature, and the finite measurement resolution (about
$\pm 5 \%$ accuracy of the reflectivity measurement \cite{refc3}). These
made the measured absorption spectrum to be broadened and led to the
slight discrepancy between the experiment result and our calculation.

In Figure 5, we plot the calculated optical-absorption coefficients versus
the rescaled photon frequency relations with $U=1.0t_{0}$ and $V=0.15t_{0}$
for different $\omega _{\pi }$ and $\alpha ^{2}/K$. Quite remarkably, when
we scale the optical-absorption coefficient by the peak value, and scale the
photon frequency by $\Gamma $, the half-width for the low photon frequency
side of the peak, we find that the scaled curves for various $\omega _{\pi }$
and $\alpha ^{2}/K$ values have an universal form below the peak photon
frequency. In their studies of temperature dependence of subgap optical
conductivity for KCP(Br) and {\it trans}-polyacetylene, J.W. Wilkins {\it et
al}. \cite{refc4} discovered that the strong subgap tail of the conductivity
had an universal scaling form. We use the same scaling method as that of
J.W. Wilkins {\it et al.}, but to study the $\omega _{\pi }$ and $\alpha
^{2}/K$ dependence of optical-absorption coefficient. Our universal form is
similar to that of J.W. Wilkins {\it et al}., which implies that the
temperature effect and nonadiabatic effect on quasi-one-dimensional Peierls
systems have similar characteristic properties.

\section{Conclusions}

The effects of quantum lattice fluctuations on the optical-absorption
coefficient in the ground state of $MX$ chain are studied by introducing an
energy-dependent e-ph scattering function $\delta(k^{\prime},k)$ via an
unitary transformation. The functional dependence of $\delta(k^{\prime},k)$
on $k^{\prime}$ and $k$ is determined by the vanishing of the second-order
perturbation of first-order terms in the transformed Hamiltonian. By using
the retarded Green's-function method we have shown that our theory gives a
good description of the effects of quantum lattice fluctuations and change
of various parameters on the optical-absorption. The inverse-square-root
singularity of optical-absorption spectrum at the gap edge in the adiabatic
case disappears because of the nonadiabatic effect. The calculated
optical-absorption coefficient is consistent well with the experiment
observation of optical conductivity measurements. The curve of subgap
optical-absorption coefficient has an universal scaling form.

\newpage {\bf {\Large Figure Caption}}\newline

{\bf Fig. 1.} The relationships between the change of optical-absorption
shape and the different phonon frequencies in the cases of
$\alpha^{2}/K=0.5t_{0}$, $U=1.0t_{0}$ and $V=0.15t_{0}$ with
$\omega_{\pi}=0.004t_{0}, 0.008t_{0}$, and $0.012t_{0}$. For comparison, the
adiabatic $(\omega_{\pi}=0)$ result is also shown which has an
inverse-square-root singularity at gap edge.\newline

{\bf Fig. 2.} The calculated optical-absorption coefficients $\alpha(\omega)$
versus the normalized photon frequency relations for $\omega_{\pi}=0.008t_{0}
$, $U=1.0t_{0}$ and $V=0.15t_{0}$ in the cases of $\alpha^{2}/K=0.3t_{0}$,
$0.4t_{0}$, and $0.5t_{0}$.\newline

{\bf Fig. 3 (a).} The calculated optical-absorption coefficients
$\alpha(\omega)$ versus the normalized photon frequency relations for
$\omega_{\pi}=0.008t_{0}$, $\alpha^{2}/K=0.5t_{0}$ and $V=0.15t_{0}$ in the
cases of $U=0.6t_{0}, 0.8t_{0}$, and $1.0t_{0}$.\newline

{\bf Fig. 3 (b).} The calculated optical-absorption coefficients
$\alpha(\omega)$ versus the normalized photon frequency relations for
$\omega_{\pi}=0.008t_{0}$, $\alpha^{2}/K=0.5t_{0}$ and $U=1.0t_{0}$ in the
cases of $V=0$, $0.15t_{0}$, $0.25t_{0}$, and $0.35t_{0}$.\newline

{\bf Fig. 4.} The rescaled optical-absorption coefficient
$\alpha(\omega)/\alpha_{{\rm peak}}$ versus the normalized photon energy
$\omega/\omega_{{\rm peak}}$ relations, where $\alpha_{{\rm peak}}$ and
$\omega_{{\rm peak}}$ are the optical-absorption coefficient and the photon
energy at the peak, respectively. The solid circles are the result of
experiment observation for
[Pt(en)$_{2}$][PtX$_{2}$(en)$_{2}$](ClO$_{4}$)$_{4}$
(Ref.\cite{refc3}, Fig.6). The solid line is our result by using
$\omega_{\pi}=0.01t_{0}$, $U=1.0t_{0}$, $V=0.3t_{0}$, and
$\alpha^{2}/K=0.2t_{0}$.\newline

{\bf Fig. 5.} Scaling plot of the calculated optical-absorption coefficients 
$\alpha(\omega)/\alpha_{{\rm peak}}$, where the photon energy is scaled by
$\Gamma$, the half-width for the low photon energy side of the peak. Note we
claim only that the subgap($\omega<\omega_{{\rm peak}}$) optical-absorption
coefficients scale.

\end{document}